\documentclass[twocolumn,twoside]{IEEEtran}
\usepackage{mathpple}
\usepackage{times}
\usepackage{amsmath}  
\usepackage{amssymb}  
\usepackage{mathrsfs} 
\usepackage{theorem}  
\usepackage{comment}  
\usepackage{upref}
\usepackage{amsfonts}
\usepackage{verbatim}
\usepackage[dvipsnames,usenames]{color}
\usepackage{enumerate}
\usepackage{graphicx}
\usepackage{subfigure}
\usepackage{latexsym}
\usepackage{mathtools} 
\usepackage{array, tabularx} 
\usepackage[shortlabels]{enumitem}
\usepackage{url}

\usepackage{times,color}
\usepackage{enumitem}
\usepackage{float}
\usepackage{listings}
\usepackage{tikz}
\usepackage{multicol}
\usepackage{psfrag}
\usepackage{pgfplots}
\usepackage{hhline}
\usepackage{xcolor,colortbl}
\usepackage{algorithm,algpseudocode,refcount} 
	\usepackage{pgf, tikz}
\usetikzlibrary{arrows, automata}

\renewcommand{\algorithmicrequire}{\textbf{Input:}}
\renewcommand{\algorithmicensure}{\textbf{Output:}}
\algnewcommand\algorithmiccase{\textbf{case}}
\algdef{SE}[CASE]{Case}{EndCase}[1]{\algorithmiccase\ #1}{\algorithmicend\ \algorithmiccase}%
\algtext*{EndCase}%
\makeatletter
\newcommand{\removelatexerror}{\let\@latex@error\@gobble}

\newcommand{\limup}[1]{\lim_{#1\rightarrow\infty}}

\newcommand{\limsupup}[1]{\limsup_{#1\rightarrow\infty}}

\newcommand{\ltuple}{${\ell\text{-tuple }}$}
\newcommand{\ltuples}{${\ell\text{-tuples }}$}
\newcommand{\vavoiding}{${\bfv\text{-avoiding }}$}

\newcommand{\be}[1]{\begin{equation}\label{#1}}
\newcommand{\ee}{\end{equation}}

\newcommand{\bc}{\begin{center}}
	\newcommand{\ec}{\end{center}}

\newcommand{\floorenv}[1]{\left\lfloor #1 \right\rfloor}

\makeatletter
\renewcommand{\thefigure}{{\@arabic\c@figure}}
\renewcommand{\fnum@figure}{{\bf Figure\,\thefigure}}
\makeatother

\newcommand{\cA}{{\cal A}}

\newcommand{\cD}{{\cal D}}
\newcommand{\cE}{{\cal E}}

\newcommand{\cO}{{\cal O}}

\newcommand{\cR}{{\cal R}}
\newcommand{\cS}{{\cal S}}

\newcommand{\bfa}{{\boldsymbol a}}

\newcommand{\bfr}{{\boldsymbol r}}
\newcommand{\bfs}{{\boldsymbol s}}

\newcommand{\bfu}{{\boldsymbol u}}
\newcommand{\bfv}{{\boldsymbol v}}
\newcommand{\bfw}{{\boldsymbol w}}
\newcommand{\bfx}{{\boldsymbol x}}

\renewcommand{\le}{\leqslant}

\renewcommand{\ge}{\geqslant}

\newcommand{\Cref}[1]{Co\-rol\-la\-ry\,\ref{#1}}

\theoremstyle{plain} \theorembodyfont{\normalfont\slshape}

\newtheorem{thm}{Theorem$\!$}
\newenvironment{theorem}{\begin{thm}\hspace*{-1ex}{\bf.}}{\end{thm}}

\newtheorem{prop}[thm]{Proposition$\!$}

\newtheorem{lem}[thm]{Lemma$\!$}
\newenvironment{lemma}{\begin{lem}\hspace*{-1ex}{\bf.}}{\end{lem}}

\newtheorem{cor}[thm]{Corollary$\!$}
\newenvironment{corollary}{\begin{cor}\hspace*{-1ex}{\bf.}}{\end{cor}}

\newtheorem{cons}[thm]{Construction$\!$}

\newtheorem{defi}[thm]{Definition$\!$}
\newenvironment{definition}{\begin{defi}\hspace*{-1ex}{\bf.}}{\end{defi}}

\newtheorem{cl}[thm]{Claim}

\theorembodyfont{\normalfont}

\newtheorem{exam}{Example$\!$}
\newenvironment{example}{\begin{exam}\hspace*{-1ex}{\bf .}}{\end{exam}}

\newtheorem{remrk}{Remark$\!$}
\newenvironment{remark}{\begin{remrk}\hspace*{-1ex}{\bf .}}{\end{remrk}}

	 \newcommand{\qed}{\hfill \mbox{\raggedright \rule{.07in}{.1in}}}

\definecolor{Codecolor}{named}{White}  

\newcommand{\Copen}{\mbox{\{\kern-5.50pt\{}}
\newcommand{\Cclose}{\mbox{\}\kern-5.50pt\}}}
\newcommand{\Cslash}{\mbox{$\backslash\kern-6.02pt\backslash$}}

\newcommand{\edge}[3]{#1\overset{#3}{\rightarrow}#2}

\newcounter{algsubstate}
\renewcommand{\thealgsubstate}{\alph{algsubstate}}

\newcommand{\red}[1]{\text{red}({#1})}

\makeatother		

\begin{document}
		
	\title{\textbf{Covering Sequences for $ \ell $-tuples}\vspace{0ex}}


\author{\large Sagi~Marcovich,~\IEEEmembership{Student Member,~IEEE}, Tuvi~Etzion,~\IEEEmembership{Fellow Member,~IEEE} and Eitan~Yaakobi,~\IEEEmembership{Senior Member,~IEEE}
	\thanks{S. Marcovich, T. Etzion and E. Yaakobi are with the Department of Computer Science, Technion --- Israel Institute of Technology, Haifa 3200003, Israel (e-mail: \texttt{\{sagimar,etzion,yaakobi\}@cs.technion.ac.il}).}}
\makeatletter
{}
\makeatother

\maketitle

\begin{abstract}	
	de Bruijn sequences of order \(\ell\), i.e., sequences that contain each \ltuple as a window exactly once, have found
	many diverse applications in information theory and most recently in DNA storage. This family of binary sequences has asymptotic rate of~$1/2$.  To overcome this low rate, we study $\ell$-tuples covering sequences, which impose that each $ \ell $-tuple appears at least once as a window in the sequence. 
	The cardinality of this family of sequences is analyzed while assuming that $ \ell $ is a function of the sequence length $ n $. Lower and upper bounds on the asymptotic rate of this family are given. 
	Moreover, we study an upper bound for $ \ell $ such that the redundancy of the set of $ \ell $-tuples covering sequences is at most a single symbol. Lastly, we present efficient encoding and decoding schemes for $ \ell $-tuples covering sequences that meet this bound.
\end{abstract}
	\maketitle

	\section{Introduction}\label{sec:intro}
	The binary de Bruijn graph of order $ \ell $, $ G_\ell $, was introduced in 1946 by de Bruijn \cite{DeB1946}.
His target in introducing this graph was to find a recursive method to enumerate the number of cyclic
binary sequences of length $2^\ell$ such that each $\ell$-tuple appears as a window exactly once in each sequence.
These sequences were later called  \emph{de Bruijn sequences}. 
These results were later generalized in~\cite{AaDb51} for any alphabet of finite size $q$, using a $ q $-ary generalization of the de Bruijn graph of order~$ \ell $, $ G_{q,\ell} $.

The vertices of $G_{q,\ell}$ are the $ q $-ary $(\ell-1)$-tuples, and its edges correspond to the $ q $-ary $\ell$-tuples.
There is an edge ${u \rightarrow  v}$ if $v$ can be obtained from $u$ by shifting one entry left and appending a symbol.
Eulerian cycles in de Bruijn graphs, i.e., cycles that visit all
the edges of $G_{q,\ell}$ exactly once, are called \emph{de Bruijn cycles}.
It was proved that the number of de Bruijn cycles in $G_{q,\ell}$ is $ (q!)^{q^{\ell-1}}/{q^\ell} $~\cite{AaDb51}. 

Each de Bruijn cycle induces a single (cyclic) de Bruijn sequence of length $ q^{\ell} $, by picking any edge in the cycle as a starting point, considering its first entry and appending the first entry of each consecutive edge in the cycle. All sequences that can be generated in this way are considered as the same sequence. Contrary to this, each de Bruijn cycle induces $q^\ell$ distinct \emph{acyclic de Bruijn sequences}, i.e., sequences of length $ q^{\ell} + \ell -1 $ that contain each $ \ell $-tuple as a window exactly once, using a similar method with the exception of appending the ($\ell-1$)-suffix of the last edge as well; each sequence corresponds to a choice of different starting edge from $ G_{q,\ell} $. Hence, the number of such acyclic
de Bruijn sequences is $(q!)^{q^{\ell-1}}$ and their asymptotic rate is $\log_q(q!) / q$ (equals $1/2$ for $q=2$). One of the first applications of the de Bruijn graph
was in the introduction of shift-register sequences and linear feedback shift registers~\cite{Go67}.
Throughout the years, an extensive number of papers have studied the de Bruijn sequences and their applications, several of those include~\cite{CHAN1982233,ComPevTes11,Fr82, Fr75, Le70,MAURER1992421,Ra81}.  Most recently, DNA storage has brought fresh interest to this family of sequences; for more information on such applications the reader is referred to~\cite{AlBrFaJa17,KiaPulMil16,SoGeGoLiYu20}.

This paper studies a novel generalization of de Bruijn sequences (for the rest of this paper we refer only to acyclic sequences). We say that a sequence is an \emph{$ \ell $-tuples covering sequence} if it contains each $ q $-ary $\ell$-tuple as a window at least once. This work follows recent generalizations of de Bruijn sequences that proposed unique variations regarding the appearances of $ \ell$-tuples in the sequence: \emph{$ \ell $-repeat free sequences}~\cite{EliGabMedYaa19IEEE, GabMil18} require each $\ell$-tuple to appear at most once, \emph{($b,\ell$)-locally-constrained de Bruijn sequences}~\cite{ChEtKiKhYa21} require each {$ \ell$-tuple} to appear at most once in every window of length $ b $, and \emph{$ (\ell, \mu)  $-balanced de Bruijn sequences}~\cite{MarEtzYaa21} require each ${\ell\text{-tuple}}$ to appear exactly $ \mu $ times in the sequence. 

Notice that for sequences of length $ q^{\ell} + \ell - 1 $, all $ \ell $-tuples covering sequences are simply the de Bruijn sequences deduced from the de Bruijn graph $ G_{q,\ell} $; as a result, their asymptotic rate is $\log_q(q!) / q$. Our main goal is to  efficiently construct codes of $ \ell $-tuples covering sequences with  higher rates (specifically larger than $1/2$ for binary sequences) and fixed number of redundancy symbols. We study the cardinality for the set of $ \ell $-tuples covering sequences and present lower bounds on its asymptotic rate for various values of $ \ell $. Additionally, we present an upper bound on $ \ell $ such that the redundancy of {the} set of {all} $ \ell $-tuples covering sequences is at most one symbol. Later, we present an encoding algorithm for the set of binary  $ \ell $-tuples covering sequences that uses a single redundancy bit and meets this bound on $ \ell $. Finally, we use a generalization of {the} de Bruijn graph to develop an upper bound for the cardinality of this set of sequences. 

Another interesting family of sequences is introduced as a building block to our analysis of $ \ell $-tuples covering sequences. For some $ \ell $-tuple $ \bfv $, we say that a sequence is a \emph{$\bfv$-avoiding sequence} if it does not contain $ \bfv $ as a window. Note that if $ \bfv $ is the all-zero $ \ell $-tuple, then this family of sequences is known as RLL sequences and was studied before, for example in~\cite{LevYaa18, schouhamer1991coding}. We study this family of sequences for any $\ell$-tuple $ \bfv $. 

The rest of this paper is organized as follows. In Section~\ref{sec:def} we formally define the families of sequences studied in this
paper and review several previous results. In Section~\ref{sec:avoid}, we study the family of $ \bfv $-avoiding sequences for any $ \ell $-tuple $ \bfv $. Based on these results, in Section~\ref{sec:tuples-covering} we analyze the cardinality of $ \ell $-tuples covering sequences and present an encoding scheme for $q=2$ which uses a single redundancy bit. 

	
	\section{Definitions and Preliminaries}\label{sec:def}
	For two integers $ i, k \in \mathbb{N} $ such that $ i \le k $ we denote by $ [i,k] $ the set $ \{i, \dots, k\} $ and use $ [k] $ as a shorthand for $  [0,k-1] $. We use the notation $ \Sigma_q = \{0,1,\dots,q-1\} $ as the alphabet of finite size $ q $. For simplicity, when $ q = 2 $, we omit the parameter~$ q $ from this notation and similar ones.

Let $ n \in \mathbb{N} $ and let $ \bfw = (w_0,\dots,w_{n-1}) \in \Sigma_q^n $ denote a sequence of length $ n $. For two positive integers $i$ and $k$ such that $ i+k-1 \le n $, let $ \bfw_{i,k} $ denote the substring $ (w_i,\dots,w_{i+k-1}) $. Additionally, let Pref$_k(\bfw) \triangleq \bfw_{0,k}, $ Suff$_k(\bfw) \triangleq \bfw_{n  - k, k} $ denote the \emph{$k$-prefix}, \emph{$k$-suffix} of $\bfw$, respectively.
The notation  $ \bfw \circ \bfv $ is the concatenation of $ \bfw $ and another sequence~$ \bfv $, and $ \bfw^i $ denotes the concatenation of $ \bfw $ $ i $ times, i.e.,  $ {\bfw^i = \bfw \circ \bfw^{i-1}}$.  
Let $ W,V $ denote two sets of vectors over $ \Sigma_q $. 
We denote the set $ WV = \{\bfw \circ \bfv \mid \bfw \in W, \bfv \in V\} $ and the set $ W^i $ to be $ i $ concatenations of the set $ W $.
Finally, the redundancy of a set $ A \subseteq \Sigma_q^n $ is defined as $ \red{A} \triangleq n - \log_q|A| $.

\begin{definition}\label{def:db-graph}
	The $ \ell $-th order $ q $-ary \textbf{de Bruijn graph} $ G_{q,\ell} $ is the digraph $ (V,E) $, where $ V = \Sigma_q^{\ell-1} $ and $$ E = \{ \left( (s_0, s_1, \dots,s_{\ell-2} ), (s_1,s_2, \dots,s_{\ell-1}) \right) \mid s_i \in \Sigma_q \}.$$
\end{definition}
Note that the edges of $G_{q,\ell}$ correspond to the set of $q$-ary ${\ell\text{-tuples}}$, $ \Sigma_q^{\ell} $.

\begin{definition}
	Let $ \ell > 1 $ be an integer and $ n = q^{\ell} + \ell - 1 $. A sequence $ \bfs \in \Sigma_q^n $ is called a \textbf{de Bruijn sequence} of order $\ell$ if $ \bfs $ contains each $q$-ary $ \ell $-tuple as a window exactly once.
\end{definition}

Let  $ \cS_q(\ell) $ denote the set of $q$-ary de Bruijn sequences of order $\ell$. 
The connection between Eulerian cycles in $G_{q,\ell}$ to de Bruijn sequences is as follows. 
In order to generate a sequence from a cycle, we pick any edge  in the cycle and set its first entry as the start of the sequence. Then, we append to the sequence the first entry of each consecutive edge in the cycle. Finally, we append the $(\ell-1)$-suffix of the last edge of the cycle to form the whole sequence. 
Note that since each edge of $ G_{q,\ell} $ can be picked as the first edge of the Eulerian cycle, a single cycle generates $ q^{\ell} $ unique de Bruijn sequences.

\begin{example}
	Let $q=2,\ell=3,n=10 $. The sequence $ \bfs~=~0001011100 $ is a de Bruijn sequence. 
	$\bfs$ can be generated from $ G_{q,\ell} $ using the Eulerian cycle 
	$$ \edge{00}{00}{000} \edge{}{01}{001} \edge{}{10}{010} \edge{}{01}{101} \edge{}{11}{011}\edge{}{11}{111} \edge{}{10}{110}\edge{}{00}{100}. $$
\end{example}

Recall that the number of de Bruijn cycles in $ G_{q,\ell} $ is $ (q!)^{q^{\ell-1}}/{q^\ell} $. Since each de Bruijn cycle generates $ q^\ell $ unique de Bruijn sequences of length $ q^\ell + \ell - 1 $, it follows that $|\cS_q(\ell)| = (q!)^{q^{\ell-1}}$. 
Therefore, the asymptotic rate of $ \cS_q(\ell) $ is 
\vspace{-2ex}
\begin{align*}  \limsupup{\ell} \frac{\log_q|\cS_q(\ell)|}{q^\ell + \ell - 1}   = \frac{\log_q(q!)}{q}. 
\end{align*}
Note that when $q=2$, this asymptotic rate equals $1/2$. However, for $q \rightarrow \infty$, it approaches $1$.

%

Next, we introduce the main family of sequences that is discussed in this paper. 
\begin{definition} \label{def:win-cover}
	Let $ n,\ell $ be integers. A sequence $\bfw\in\Sigma_q^n$ is called an \textbf{$\ell$-tuples covering sequence} if $ \bfw $ contains each $ q $-ary $ \ell $-tuple as a window at least once, i.e., for each $ \bfv \in \Sigma_q^\ell $, there exists $ i \in [n-\ell+1] $ such that $ \bfw_{i,\ell} =  \bfv$.
\end{definition}
\begin{example}
	Let $ q=2,\ell=3,n=13 $. The sequence $ \bfw_1 = 0001001110101 $ is an $ \ell $-tuples covering sequence. However, the sequence $ \bfw_2 = 1001001110101 $ is not an $ \ell $-tuples covering sequence, since it does not contain the $ 3 $-tuple $ 000 $.
\end{example}

We denote the set of all $q$-ary $ \ell $-tuples covering sequence over $ \Sigma_q^n $ by $ \cR_{q}(n,\ell) $ and notate the size of such code by $ r_q(n,\ell) \triangleq |\cR_q(n,\ell)| $. For a window length that is a function of $ n$, that is $ \ell = f(n) $, we denote the asymptotic rate of $ \cR_{q}(n,f(n)) $ by $$ \mathbb{R}_q(\ell) \triangleq \limsupup{n} \frac{\log_q r_q(n,f(n))}{n}.  $$ 

Note the following connection between $\ell$-tuples covering sequences and de Bruijn sequences; if $ n = q^\ell + \ell - 1 $, then the set $ \cR_q(n,\ell) $ is exactly the set of de Bruijn sequences $ \cS_q(\ell) $.  Therefore, $ \mathbb{R}_q(\ell) = {\log_q(q!)}/{q} $ in this case.  In Section~\ref{sec:tuples-covering} we study the cardinality of $ \cR_{q}(n,\ell) $ for various sizes of $ \ell $, i.e., for various functions $f(n)$. Moreover, we present an encoding algorithm for $q=2$ that uses a single redundancy bit.

%
	
	\section{$\bfv$-Avoiding Sequences}\label{sec:avoid}
	In this section, we present the auxiliary family of \emph{$\bfv$-avoiding sequences} that is used later in our analysis of $\ell$-tuples covering sequences in Section~\ref{sec:tuples-covering}. 

\begin{definition}\label{def:non-present}
	Let $\ell$ be an integer and $ \bfv \in \Sigma_q^\ell $ a fixed $\ell$-tuple. The set of \textbf{$ \bfv $-avoiding sequences} over $ \Sigma_q^n $, denoted by $ \cA_q(n,\bfv) $ contains all  $q$-ary sequences of length $n$ that do not contain $ \bfv $ as a window.  Namely,
	$$ \cA_q(n,\bfv) = \{ \bfw \in \Sigma_q^n \mid \forall i \in [n-\ell+1], \bfw_{i,\ell}\neq\bfv \} $$
\end{definition}
For a given $ \ell $-tuple $ \bfv $, we notate the size of this code by $ a_q(n,\bfv) \triangleq | \cA_q(n,\bfv) | $.
Note that for $\bfv = 0^{\ell}$, this family of sequences is the family of $(0,\ell-1)$-RLL sequences~\cite{schouhamer1991coding} (for integers $d,k$, a $(d,k)$-RLL sequence satisfies that the number of zeros between two consecutive ones is in the range $[d,k]$). These sequences were studied extensively in~\cite{LevYaa18} for different functions $\ell=f(n)$. 

We are motivated to study this family of sequences due to the following connection to the family of $\ell$-tuples covering sequences; a sequence~$\bfs $ is an $\ell$-tuples covering sequence if and only if for every $ \bfv \in \Sigma_q^\ell $, $ \bfs $ is not a $\bfv$-avoiding sequence. This connection will be utilized later in order to encode and analyze the cardinality of $ \cR_q(n,\ell) $. 

\subsection{Cardinality Upper Bound Analysis}\label{sec:avoid-card}

First, we give an upper bound for $a_q(n,\bfv)$ for any $ \bfv \in \Sigma^\ell_q $ in order to use it later to estimate the cardinality of $ \cR_{q}(n,\ell) $.
For a sequence $ \bfs \in \Sigma_q^n $, let $ p(\bfs) $ denote its \emph{period}, that is, the smallest positive integer that satisfies $ s_{i} = s_{i+p(\bfs)} $ for every $ i \in [n-p(\bfs)] $. Initially, we prove the following useful lemma. 
\begin{lemma}\label{clm:help-beta}
	Let $ \ell $ be an integer and $ \bfv \in \Sigma_q^\ell $ a tuple of length $\ell$. Let $ \beta(\bfv) $ denote the set of sequences of length $ 2\ell $ that contain $ \bfv $ as a window at least once, i.e., 
$$ \beta(\bfv) =  \Sigma_q^{2\ell} \setminus \cA_{q}(2\ell,\bfv). $$
 Then, $$ |\beta(\bfv)| \ge \frac{1}{2}(\ell+1)(q-1)^2q^{\ell-2}. $$
\end{lemma}
\begin{IEEEproof}
	We observe that $ \bfv $ can appear in a  sequence of length $ 2\ell $ many times only if its period is small. In particular, we show that if $ p(\bfv) > \ell/2 $ then $ \bfv $ can appear at most twice. Hence, we lower bound $ |\beta(\bfv)| $ by placing $ \bfv $ in a sequence $ \bfx \in \beta(\bfv)$ and setting {the} minimal number of symbols in order to ensure that $ \bfv $ appears at most twice in $ \bfx $. At last, we account for occurrences that were enumerated multiple times. 
	
	Let $ i \in [\ell+1] $ denote the starting position of $ \bfv $ in $ \bfx $, and assume w.l.o.g that $ \bfv $ is in the middle of $ \bfx $, i.e., $ i \in [1,\ell]  $. We increase the period of $ \bfx $ forward by selecting $ x_{i+\ell} \neq v_{\ell \bmod p(\bfv)} $ in order to ensure that $p(\bfx_{i,\ell+1}) > \ell/2$. Let $p = p(\bfx_{i,\ell+1})$ and assume in the contrary that $ p \le \ell/2 $. However, this means that $ p $ is a multiple of $ p(\bfv) $ and hence since $ x_{i+\ell} \neq v_{\ell \bmod p(\bfv)}, $ then $ x_{i+\ell} \neq x_{i+\ell-p} $ which is a contradiction to the definition of $ p $. 
	Similarly, we increase the period of $ \bfx $ backwards by selecting  $ x_{i-1} \neq v_{(-1 \bmod p(\bfv))} $. If $ \bfx $ starts or ends with $ \bfv $, we only need to select a single symbol in order to increase the period of $ \bfv $ forward or backwards, respectively. Hence, we enumerate for each position $ i \in [\ell+1] $ at least $ (q-1)^2q^{\ell-2} $ possible choices of $ \bfx $.
	
	Next, it is clear that another instance of $ \bfv $ can appear at most once after $ i $, at position $ i' > i + \ell/2 $, and once before $ i $, at position $ i'' < i -  \ell/2 $. Moreover, it is only possible for one of those instances to occur, since $ |[i'', i' + \ell]| > 2\ell $. Thus, when summing the possible choices of $ \bfx $ for every possible position of the tuple $ \bfv $, we count each choice at most twice since $ \bfv $ can appear at most twice in $ \bfx $. Thus, 
	$$ |\beta(\bfv)| \ge 
	\frac{1}{2}(\ell+1)(q-1)^2q^{\ell-2}. $$	
\end{IEEEproof} 	

Next, we use the result of Lemma~\ref{clm:help-beta} to obtain an upper bound on $ a_q(n,\bfv) $ for every $ n, \ell $ and $ \bfv \in \Sigma_q^{\ell} $.
\begin{lemma}\label{lem:aq-nkv}
	Let $ n,\ell $ be positive integers such that $ \ell \le n $, and let $ \bfv \in \Sigma_q^\ell $ be any sequence of length $ \ell $. Then, $$ a_{q}(n,\bfv) \le q^{n-c_1\frac{n-2\ell}{q^\ell}},  $$ where $ c_1 = \frac{(q-1)^2\log_q e}{4q^2} $ and $ e $ is the base of the natural logarithm.
\end{lemma}
\begin{IEEEproof}
	Consider the following set $$ \cA'_q(n,\bfv) = \cA_q(2\ell,\bfv)^{\floorenv{\frac{n}{2\ell}}} \Sigma_q^{n \bmod 2\ell}, $$ 
	which is the set of sequences which are concatenation of $ \floorenv{\frac{n}{2\ell}} $ sequences from $\cA_q(2\ell,\bfv) $ appended by any  sequence of length $ n \bmod 2\ell $.	
	Note that $ \cA_q(n,\bfv) \subseteq \cA'_q(n,\bfv) $ since for every $ \bfx \in \cA_q(n,\bfv) $ and $ i \in [\floorenv{\frac{n}{2\ell}}] $, $\bfv $ is not a substring of $ \bfx_{2\ell\cdot i,2\ell} $, i.e., $\bfx_{2\ell\cdot i,2\ell} \in \cA_q(2\ell,\bfv) $. Hence,
	\begin{align}\label{eq:aq-nkv1}
		a_q(n,\bfv) \le |\cA'_q(n,\bfv)| = a_q(2\ell,\bfv)^{\floorenv{\frac{n}{2\ell}}} q^{n \bmod 2\ell}.
	\end{align}
	
	All sequences of length $2\ell$ that contain $ \bfv $ at least once are not included in $ \cA_q(2\ell,\bfv) $. Therefore, using Lemma~\ref{clm:help-beta},
	\begin{equation}\label{eq:aq-nkv2}
		\begin{aligned}
			a_q(2\ell,\bfv)= q^{2\ell} - |\beta(\bfv)| &\le q^{2\ell} - 	\frac{1}{2}(\ell+1)(q-1)^2q^{\ell-2} 
			\\&=  q^{2\ell} \left(1 - \frac{(\ell+1)(q-1)^2}{2q^{\ell+2}}\right).
		\end{aligned}
	\end{equation}
	By combining inequalities (\ref{eq:aq-nkv1}) and (\ref{eq:aq-nkv2}) we get
	\begin{align*}
		a_q(n,\bfv) &\le \left( q^{2\ell} \left( 1 - \frac{(\ell+1)(q-1)^2}{2q^{\ell+2}}\right)\right)^{\floorenv{\frac{n}{2\ell}}}\cdot q^{n \bmod 2\ell}
		\\&=q^n \left( 1 - \frac{(\ell+1)(q-1)^2}{2q^{\ell+2}}\right)^{\floorenv{\frac{n}{2\ell}}}
		\\&\overset{(a)}{\le}q^n \cdot \exp\left(- \frac{(\ell+1)(q-1)^2}{2q^{\ell+2}}\right)^{\floorenv{\frac{n}{2\ell}}}
		\\&\le q^n \cdot \exp\left(- \frac{(\ell+1)(q-1)^2}{2q^{\ell+2}} (\frac{n}{2\ell}-1)\right)
		\\&\le q^{n - \frac{(q-1)^2(n-2\ell)}{4q^{\ell+2}}\log_q e}
	\end{align*}
	where $ (a) $ results from the inequality $ 1 - x \le e^{-x} $ for all $ x $.
\end{IEEEproof}

\subsection{Compression Algorithm for Binary Sequences}\label{sec:avoid-comp}
Next, we focus on binary sequences, i.e., $q=2$, and present  a \vavoiding sequences compression algorithm for any $ \bfv $ of length $ \ell \le \log n - 6 $ and $ n $ large enough. The algorithm receives a $ \bfv $-avoiding sequence of length $n$ and outputs a unique unconstrained sequence of length $n-1$. Clearly, this algorithm can be used for any $ \ell' < \ell $ by padding $ \bfv $ to size $ \ell $ and continuing regularly; hence we assume from now on that $ \ell = \log n - 6 $. This compression algorithm will be utilized in Section~\ref{sec:tuples-covering} to encode binary $\ell$-tuples covering sequences. 

For every $ \bfv \in \Sigma^{\ell} $, we denote two functions, 
$$ f_1(\bfv) =  \bfv \circ (1 - v_{|\bfv| \bmod p(\bfv)}) $$
$$ f_2(\bfv) =  \text{Pref}_{\lfloor |\bfv|/2 \rfloor + 3 }(\bfv) \circ f_1(\text{Suff}_{\lceil |\bfv|/2 \rceil - 3 }(\bfv)). $$
Note that both functions append a single bit to $ \bfv $. We have the following lemma,
%

\begin{lemma}\label{cor:p-half}
	For every $ \bfv \in \Sigma^\ell $, $ p(f_1(\bfv)) \ge \lceil (\ell+1)/2 \rceil  $.
\end{lemma}
We say that a sequence has a \emph{long period} if its period is at least half {of} its length, i.e., $ p(\bfv) \ge \lceil |\bfv| /2 \rceil $. Hence, from Lemma~\ref{cor:p-half}, for every $ \bfv \in \Sigma^\ell $, $ f_1(\bfv) $ has a long period, and $f_2(\bfv)$ satisfies that its $ (\lceil \ell/2 \rceil - 2) $-suffix has a long period. These functions are utilized in the following compression algorithm.

\begin{figure}[h]
	\rule[1ex]{\columnwidth}{0.1pt}
	\includegraphics[width=0.9\columnwidth] {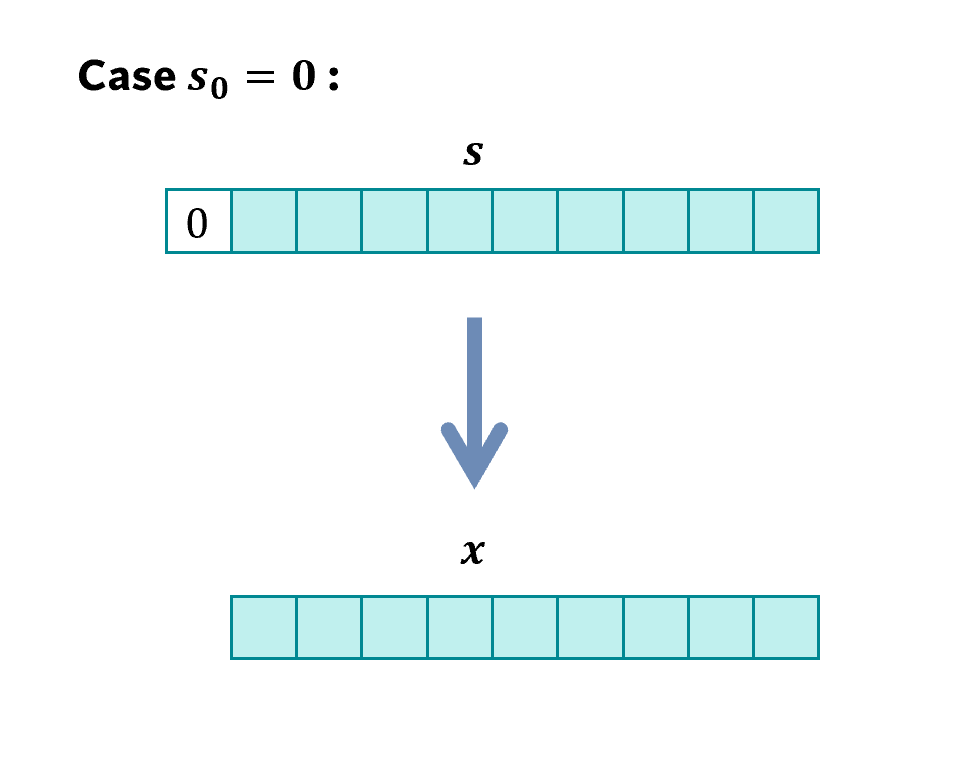}
	\caption{Illustration of Algorithm~\ref{alg:compress} for the case $s_0 = 0$.}
	\label{fig:nis1}
	\rule[1ex]{\columnwidth}{0.1pt}
\end{figure}

\begin{figure}[h]
	\rule[1ex]{\columnwidth}{0.1pt}
	\includegraphics[width=0.9\columnwidth] {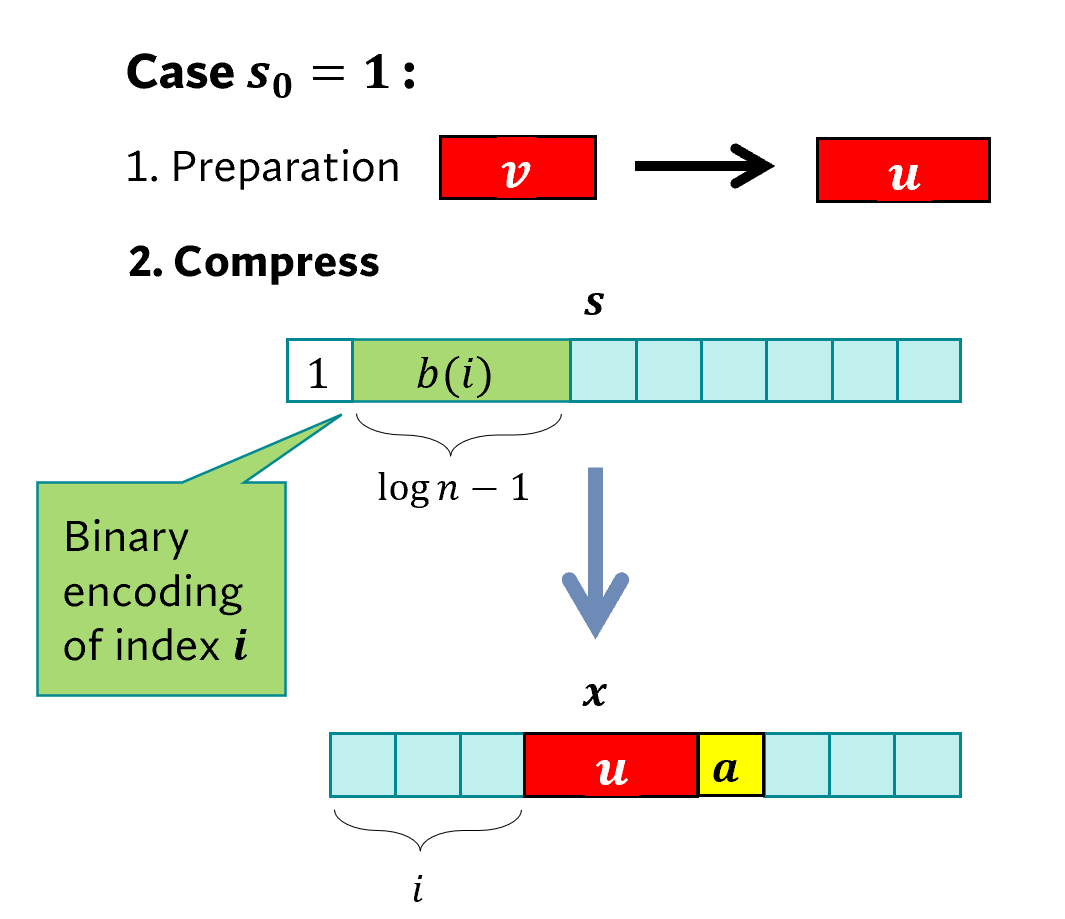}
	\caption{{Illustration of Algorithm~\ref{alg:compress} for the case $s_0 = 1$.}}
	\label{fig:nis2}
	\rule[1ex]{\columnwidth}{0.1pt}
\end{figure}

The $\bfv$-avoiding compression algorithm (Algorithm~\ref{alg:compress}) receives a sequence $ \bfs \in \cA(n,\bfv) $ for $ \bfv \in \Sigma^{\ell} $ and compresses it to some uniquely decodable  sequence $ \bfx \in \Sigma^{n-1} $. 
Initially, the algorithm checks the first bit of $ \bfs $. If it is zero, then the rest of $ \bfs $ is returned as the result (see Figure~\ref{fig:nis1}). Otherwise, an index $ i $ is decoded from the subsequent $ \log n - 1 $ bits of $ \bfs $ (by converting this binary sequence to its integer representation) and the algorithm will construct $ \bfx $ by inserting an occurrence of $ \bfv $ at this index. However, since such an insertion might create new instances of $ \bfv $ in the sequence $ \bfx $,  $ 5 $ additional bits are appended to $ \bfv $ in order to ensure that the insertion index can always be deduced by the decoder (see Figure~\ref{fig:nis2}).
The redundancy bits are added as follows; first, two bits are appended to~$ \bfv $ (independently of the input sequence $ \bfs $) to construct $ \bfu $, a sequence that satisfies that both $ \bfu $ and  $ \text{Suff}_{\lceil (\ell+1)/2 \rceil - 2}(\bfu)$ have long periods. As a result, when $ \bfu $ is inserted at position~$ i $, at most three new occurrences can be created to the right of it (see Lemma~\ref{lem:unique-suff} which follows). These cases are eliminated using the $ 3 $ remaining bits appended to $ \bfu $, denoted by $\bfa$. The result is a sequence $ \bfx \in \Sigma^{n-1} $ with its rightmost occurrence of $ \bfu $ at position $ i $. 

\begin{algorithm}[t]
	
	\caption{$\bfv$-avoiding compression algorithm}\label{alg:compress}
	\algorithmicrequire{ A sequence $ \bfs \in \cA(n,\bfv) $} \\
	\algorithmicensure { A sequence $ \bfx \in \Sigma^{n-1} $}
	\begin{algorithmic}[1]
		\State{If $s_0=0$, return $ \bfs_{1,n-1} $. Otherwise, decode index $ i $ from $\bfs_{1,\log n-1} $ and set $ \bfw = \text{Suff}_{n-\log n}(\bfs) $}\label{step:comp-init}			
		\State{Construct $ \bfu = f_2(f_1(\bfv)) $}\label{step:comp-construct2}
		\State{Let $$ A = \left\{ p(\bfu) - 3 \le m \le |\bfu|-1 : \bfw_{i+1,m} = \text{Suff}_m(\bfu) \right\}  $$} \label{step:A}
		\State{Set $ \bfa = 000 $, notate the elements of $A$ in decreasing order $ m_2 > m_1 > m_0 \in A  $ (starting from $ m_2 $)} 
		\For{every $m_k \in A $}\label{step:comp-find}
		\State{Set $ a_{k} = 1-{u}_{\ell-1-m_k+k}$} \label{step:comp-a_k}
		\EndFor
		\State{Return $\bfx = \text{Pref}_i(\bfw) \circ \bfu \circ \bfa  \circ \text{Suff}_{|\bfw|-i}(\bfw)$}\label{step:comp-insert}
	\end{algorithmic}
\end{algorithm}


\begin{lemma}\label{lem:unique-suff}
	At Step~\ref{step:A} of Algorithm~\ref{alg:compress}, we have that $ |A| \le 3$ .
\end{lemma}
\begin{IEEEproof}
	Remind that $$ A = \left\{ p(\bfu) - 3 \le m \le |\bfu|-1 : \bfw_{i+1,m} = \text{Suff}_m(\bfu) \right\}.  $$
	For any two values $ m_0 < m_1 \in A $, it holds that $ \bfu_{\ell-m_0+1,m_0} = \bfu_{\ell-m_1+1,m_0} $ and thus $ p(\text{Suff}_{m_1}(\bfu)) \le m_1-m_0 $. Let $ \hat{\bfu} $ denote the $ (\lceil (\ell+1)/2 \rceil - 2) $-suffix of $ \bfu $ which its period was increased by invoking $ f_2 $.
	Since
	$$ m_1 > m_0 \ge p(\bfu) - 3 \ge p(f_1(\bfv)) - 3 \ge \lceil (\ell+1)/2 \rceil - 3, $$ it follows that $ \text{Suff}_{m_1}(\bfu) $ contains $ \hat{\bfu} $ and hence  $$ m_1 - m_0 \ge p(\hat{\bfu}) \overset{(a)}{\ge} \frac{\lceil (\ell+1)/2 \rceil - 2}{2} $$
	where (a) follows from Corollary~\ref{cor:p-half}. Finally, assume in the contrary that there exist four unique values $ m_0 <m_1<m_2 < m_3$ that satisfy the condition of the lemma. We have
	$$ m_3 - m_0 \ge 3(m_1-m_0) \ge  \frac{3(\lceil (\ell+1)/2 \rceil - 2)}{2}, $$
	which is larger than the difference between the maximal and the minimal values in $ [p(\bfu)-3, |\bfu|-1] $ for $ \ell $ large enough, that is,
	\begin{align*}
		|\bfu|-1 - p(\bfu) + 3 &\le \ell + 1 - \lceil (\ell+1)/2 \rceil + 3 \\&= \lfloor (\ell+1)/2 \rfloor + 3,
	\end{align*}
	and we have a contradiction.
\end{IEEEproof}

\begin{lemma}
	The assignments at Steps~\ref{step:comp-a_k} and~\ref{step:comp-insert} of Algorithm~\ref{alg:compress} are correctly defined for $ n $ large enough.
\end{lemma}
\begin{IEEEproof}
Assume in the contrary that the assignment at Step~\ref{step:comp-a_k} is incorrect for some $ m_k $, i.e., $ \ell-1-m_k+k < 0 $. It follows that $ m_k > \ell - 1 $. However, from the proof of Lemma~\ref{lem:unique-suff} $ m_k $ must be the maximal value in $ A $ and since the values of $ A $ are indexed in decreasing order, the index of $ m_k $ is $ k = 2 $ and the assignment is correctly defined. 
The assignment at Step~\ref{step:comp-insert} is correctly defined since for $ n $ large enough, {it holds that} $$ |\bfw| = n - \log n > \frac{n}{2} \ge i. $$
\end{IEEEproof}

\begin{lemma}\label{lem:insertion-index}
	After Step~\ref{step:comp-insert} of Algorithm~\ref{alg:compress}, $ \bfx $ has its rightmost occurrence of $ \bfu $ at position $ i $.
\end{lemma}
\begin{IEEEproof}
	Assume in the contrary that $ \bfx $ contains another instance of $ \bfu $ at position $ j > i $.
	Clearly, $ \bfw $ does not contain $ \bfu $ since $ \bfw $ is a substring of the input sequence $ \bfs \in \cA(n,\bfv) $. Therefore, from the construction of $ \bfx $ at Step~\ref{step:comp-insert}, $ j-i \le |\bfu|+2 $. On the other hand, $ j - i \ge p(\bfu) $. It follows that $ \bfw_{i+1,m} = \text{Suff}_m(\bfu) $ for $ m_k = j - i - 3 $ and hence $ m_k $ belongs to the set $ A $ that is constructed at Step~\ref{step:A}. Combining with Lemma~\ref{lem:unique-suff}, this occurrence was eliminated at Step~\ref{step:comp-find} of the algorithm using $$ x_{j+\ell-1-m_k+k} = x_{i+\ell+2+k} = a_k \neq u_{\ell-1-m_k+k}, $$ and we have a contradiction.
\end{IEEEproof}

\begin{theorem}\label{th:alg1}
	Algorithm~\ref{alg:compress} compresses a sequence $ \bfs \in \cA(n,\bfv) $ to a sequence $ \bfx \in \Sigma^{n-1} $ that can be uniquely decoded to its input sequence $ \bfs $. The time complexity of the Algorithm~\ref{alg:compress} is $\cO( \log n)$ and the time complexity of its decoder is $\cO(n)$. 
\end{theorem}
\begin{IEEEproof}
Clearly, if $ s_0 = 0 $ Algorithm~\ref{alg:compress} returns a sequence of length $n-1$.
Otherwise, the algorithm inserts $ \ell + 5 = \log n  - 1 $ bits at Step~\ref{step:comp-insert} after removing $ \log(n) $ bits at Step~\ref{step:comp-init}, and the length of the result is  $ n-1 $ as well. The time complexity of the algorithm follows from $ \Theta(\log n) $ to decode the index $i$, and other operations are on $\bfv$ and a substring of length $ \cO(\ell) $ of $ \bfs $. Note that the time complexity of finding the period of $\bfv$ (or a substring of it) is also $ \cO(\ell) $~\cite{Czumaj2000OnTC}. Hence, the total time complexity is $ \cO(\log n) $.

A decoder for this algorithm constructs $ \bfu = f_2(f_1(\bfv)) $ and looks for the rightmost occurrence of $ \bfu $ in $ \bfx $. If such occurrence is found, then it must be at the insertion index $ i $ from Lemma~\ref{lem:insertion-index}. From here, reconstructing $ \bfs $ is straightforward. Since the decoder scans the sequence $ \bfx $ in order to find the rightmost occurrence of $\bfv$, its time complexity is $ \cO(n) $, {which is} larger than the time complexity of the encoder.

\end{IEEEproof}

\begin{remark}
Note that Algorithm~\ref{alg:compress} is generic and fits any $ {\bfv \in \Sigma^\ell} $. Clearly, with some knowledge of the tuple $ \bfv $ some steps can be skipped and the supported tuple length $ \ell $ can be larger. For example, if $ \bfv $ has a long period it is unnecessary to invoke $ f_1 $ at Step~\ref{step:comp-construct2} and $ \ell = \log n - 5 $ can be used. If $ p(\bfv) = \ell $, i.e., $ \bfv $ is aperiodic, then no additional bits are necessary (the algorithm returns $\text{Pref}_i(\bfw) \circ \bfv \circ \text{Suff}_{|\bfw|-i}(\bfw)$) and the algorithm is applicable to $ \ell = \log n - 1 $. 
\end{remark}

\begin{example}
Let $ \ell = 9 $, $ \bfv = 010101010 $. Notice that $ p(\bfv)~=~2 $. We can construct {in} advance $ f_1(\bfv) = 0101010100, $
and since $ |f_1(\bfv)|/2 - 3 = 2 $, 
$$ \bfu = f_2(f_1(\bfv)) = 01010101  \circ f_1(00) = 01010101001. $$
Notice that $ p(\bfu) = 9 $.

Let $ n = 2^{15} $ and assume the input sequence $ \bfs_1 = 0^{n}~\in~\cA(n,\bfv)$. In this case, the algorithm simply returns $ \bfx_1 =  0^{n-1} $ at Step~\ref{step:comp-init}. The decoder receives $ \bfx_1 $ and notices that no occurrences of $ \bfv $ exist in the sequence, {and thus it} correctly returns $ 0 \circ \bfx_1 = \bfs_1 $.

Next, assume $ \bfs_2 = (10101001)^{2^{12}} \in \cA(n,\bfv) $. One can easily verify that $ |\bfs_2| = n $ and that $ \bfv $ does not appear as a window in $ \bfs_2 $. At Step~\ref{step:comp-init}, since the first bit of the sequence is~$ 1 $, the algorithm decodes an index $ i $ from the integer value of $ (\bfs_2)_{1,14} = 01010011010100,  $
that is, $ i = 5332 $, and denotes by $ \bfw $ the unused part of $ \bfs_2 $. Soon, the algorithm will insert the tuple $ \bfu $ at position $ i $. However, we need to ensure that this insertion does not create additional occurrences of $\bfu $ that start to the right of $ i $.

Since $\bfw_{i+1,10} = 1010100110 $, the set of indices that is constructed at Step~\ref{step:A} is $ A = \{10\} $. Note that we have $ |A| = 1 $ although by Lemma~\ref{lem:unique-suff} the maximal size of $ A $ is $ 3 $. At Step~\ref{step:comp-a_k} the algorithm constructs $ \bfa = 001 $ where $ a_2 = 1 $ ensures that a new occurrence of $ \bfu $ is not created when concatenating $ \bfu $ and $ \bfw_{i+1,10} $. Finally,  at Step~\ref{step:comp-insert} the algorithm returns the sequence $$ \bfx_2 = \text{Pref}_i(\bfw) \circ 01010101001 \circ 001  \circ \text{Suff}_{|\bfw|-i}(\bfw). $$

The decoder receives $ \bfx_2 $ and identifies the rightmost occurrence of $ \bfv $ at position $ i = 5332 $. Thus, it constructs $$ \bfs_2 = 1 \circ b(i) \circ \text{Pref}_i(\bfx_2) \circ \text{Suff}_{|\bfx_2|-i-5}(\bfx_2). $$	
\end{example}

	\section{Tuples Covering Sequences }\label{sec:tuples-covering}
	In this section, we focus on the main family of sequences discussed in this paper, $\ell$-tuples covering sequences. We study the cardinality of this set of sequences and present a lower bound on its asymptotic rate for various values of $ \ell $. Then, we present an upper bound for $ \ell $ such that the redundancy of {the} set of $ \ell $-tuples covering sequences is at most one symbol. Later, we present an encoding algorithm for  binary  $\ell$-tuples covering sequences that uses a single redundancy bit that meets this bound on $ \ell $. Finally, we use a generalization of {the} de Bruijn graph to develop an upper bound for the cardinality of this set of sequences. 

\subsection{{Lower Bound Analysis on the Rate}}
We begin the discussion of $ \ell $-tuples covering sequences for any alphabet of size $ q  $. For integers $ n, \ell $, it is clear that if $ n<q^\ell + \ell - 1 $ then $ r_q(n,\ell) = 0$ since a sequence of such length can contain at most $ n-\ell+1 $ unique $ \ell $-tuples where $ n-\ell+1 < q^\ell $. In the case where $ n = q^\ell + \ell - 1 $ the set $ \cR_q(n,\ell) $ is exactly the set of de Bruijn sequences of order $\ell$, and hence $r_q(n,\ell) = |\cS_q(\ell)|$. For larger values of $ n $, we have the following lemma.
\vspace{-1ex}
\begin{lemma}\label{rqnl-size}
	Let $ n = q^\ell + \ell - 1 + k $ for  $ \ell, k \in \mathbb{N} $. Then, 
	$$ r_q(n,\ell) \ge (q!)^{q^{\ell-1}} \cdot q^k. $$
\end{lemma}
\begin{IEEEproof}
	We construct $ \ell $-tuples covering sequences of length~$ n $ using any de Bruijn sequence of order $\ell$ followed by any~$ k $ symbols from $ \Sigma_q $. The result follows immediately from the size of $\cS_q(\ell)$.
\end{IEEEproof}

Therefore, we have the next results.

\begin{corollary}\label{clm:rate-lower-1}
	Let $ n = q^\ell + \ell - 1 + f(n) $ for  $ \ell \in \mathbb{N} $. Then,
	$$ \mathbb{R}_q(\ell) \ge \begin{cases}
		\frac{\log_q(q!)}{q} & f(n) = o(q^{\ell})\\
		\frac{q^{-1}\log_q(q!)+\alpha}{1+\alpha} & f(n) = \alpha q^{\ell} + o(q^{\ell}) \text{ for } \alpha  > 0 \\
		1 & f(n) = \omega(q^{\ell})\\
	\end{cases}  $$
\end{corollary}
\begin{IEEEproof}
	Using Lemma~\ref{rqnl-size} we have 
	$$ \mathbb{R}_q(\ell) \ge \limsupup{n}\frac{q^{\ell-1}\log_q(q!) + f(n)}{q^{\ell}+f(n)}, $$
	and the results for $f(n) = o(q^{\ell})$ and $f(n) = \omega(q^{\ell})$ follow immediately. 
	For the last case, we apply $ f(n) = c\cdot q^{\ell} + o(q^{\ell}) $ and receive
	$$ \mathbb{R}_q(\ell) \ge \limsupup{n} \frac{q^{\ell-1}\log_q(q!) +c q^{\ell}}{q^{\ell} + cq^{\ell}}  = \frac{q^{-1}\log_q(q!)+c}{1+c}. $$
\end{IEEEproof}

Alternatively, we have
\begin{corollary}\label{clm:rate-lower}
	Let $ n \in \mathbb{N} $ and let $ \ell = \log n - g(n) $. Then,
	$$ \mathbb{R}_q(\ell) \ge \begin{cases}
		\frac{\log_q(q!)}{q} & g(n) = o(1)\\
		1 + \frac{1}{q^{c+1}}\log_q(q!) - \frac{1}{q^c} & g(n) = c \text{ for } c > 0 \\
		1 & g(n) = \omega(1)\\
	\end{cases}  $$
\end{corollary}
\begin{IEEEproof}
	By applying $ n = q^\ell + \ell - 1 + k $, we have $ q^\ell = \frac{n}{q^{g(n)}} $, and $ k = n -\frac{n}{q^{g(n)}} -\cO(\log_q(n)) $. Thus, using Claim~\ref{rqnl-size} we have 
	\begin{align*}
	\mathbb{R}_q(\ell) &\ge \limup{n}\frac{\frac{n}{q^{g(n)+1}}\log_q(q!) + n - \frac{n}{q^{g(n)}}}{n} \\&= \limup{n} 1 + \frac{1}{q^{g(n)+1}}\log_q(q!) - \frac{1}{q^{g(n)}},	
\end{align*}
	and the result follows as $ \limup{n} \frac{1}{q^{g(n)}} = 1 $  when $ g(n) = o(1) $ and $ \limup{n}\frac{1}{q^{g(n)}} = 0 $  when $ g(n) = \omega(1) $. 
\end{IEEEproof}
For convenience, we use both representations of $ n $ and $ \ell $ throughout this section.

\subsection{{Single Symbol Analysis on the Redundancy}}

Next, we present an upper bound on $ \ell $, where the redundancy of $\cR_q(n,\ell)$ is at most a single symbol. This bound uses the upper bound on the cardinality of $ \bfv $-avoiding sequences presented in Section~\ref{sec:avoid}.
\begin{theorem}\label{lem:red-unionbound}
	If $ n, \ell $ are integers such that $ \ell \le \log_q n - \log_q \log_q n - \cO(1) $, then for $ n $ large enough, 
	$ \textmd{red}(\cR_q(n,\ell)) \le 1 $.
\end{theorem}
\begin{IEEEproof}
	Let $ \ell = \log_q(n) - \log_q(\log_q(n)) - c_2 $ for some constant $ c_2 > 0 $ that will be determined later. For every sequence $ \bfw \in \Sigma_q^n $ that is not an $\ell$-tuples covering sequence, there exists $ \bfv \in \Sigma_q^\ell $ such that $ \bfw $ is a $ \bfv $-avoiding sequence, i.e., $ \bfw \in \cA_q(n,\bfv) $. Thus, using Lemma~\ref{lem:aq-nkv}, the number of sequences that are not  $\ell$-tuples covering sequences can be bounded  above by
	\begin{equation}\label{eq:red1}
		\sum_{\bfv \in \Sigma_q^\ell} a_q(n,\bfv) \le q^{n-c_1\frac{n-2\ell}{q^\ell}+\ell}, 
	\end{equation}
	where $  c_1 = \frac{(q-1)^2\log_q e}{4q^2}  $. Therefore, in order to have $ \text{red}(\cR_q(n,\ell)) \le 1 $, i.e., $ r_q(n,\ell) \ge q^{n-1} $, we require that the value in equation~(\ref{eq:red1}) is bounded from above by $ (q-1)q^{n-1} $. Hence, it is enough to have 
	\begin{equation}\label{eq:red2}
		c_1\frac{n-2\ell}{q^\ell}-\ell \ge \log_q \frac{q}{q-1}.
	\end{equation}
	By applying $ \ell = \log_q n - \log_q\log_q n - c_2 $, or alternatively $ q^{\ell} = \frac{n}{q^{c_2}\log n} $, we get
	\begin{align*}
		c_1\frac{n-2\ell}{q^\ell}-\ell &=
		c_1\frac{n}{q^{\ell}} - \ell\left(\frac{2c_1}{q^{\ell}}-1\right)
		\\&= c_1q^{c_2}\log_q n - \ell\left(\frac{2c_1q^{c_2}\log_q n}{n}-1\right)
		\\&\ge c_1q^{c_2}\log_q n - \frac{2c_1q^{c_2}(\log_q n)^2}{n}-\log_q n
	\end{align*}
	which satisfies inequality~(\ref{eq:red2}) when $ c_1q^{c_2}  > 1 $, i.e., $ c_2 > -\log_q c_1 $, and for $ n $ large enough.
\end{IEEEproof}

\subsection{Encoding Algorithm for the Binary Case}
For the rest of this section, we assume that $q=2$. We present an encoder for $\ell$-tuples covering sequences over $ \Sigma^n $. This encoder uses a single redundancy bit and handles $\ell$-tuples of length $ \ell \le \log n - \log \log n - 6 $ for $ n $ large enough. Note that this value of $ \ell $ is associated with the bound presented in Theorem~\ref{lem:red-unionbound}.

This algorithm is based on the compressor of $\bfv$-avoiding sequences presented in Algorithm~\ref{alg:compress}. For $ \bfv \in \Sigma^{\ell} $, let $  \cE_{\bfv} $ denote this compressor, i.e., $\cE_{\bfv}$ receives a $ \bfv $-avoiding sequence of length $n$  such that $\ell \le \log n - 6$ and outputs an unconstrained and uniquely decodable sequence over $ \Sigma^{n-1} $. Let $ n_{\cE} $ denote the maximal sequence length that can be compressed with $\cE_{\bfv}$ such that $|\bfv| = \ell$, that is, $ {n_{\cE} = 2^{\ell+6} = n/\log n} $. Moreover, $ \cE_{\bfv} $ can be used to efficiently compress $ \bfv $-avoiding sequences of length $ n \ge n_{\cE} $ as well; the input sequence is split to consecutive segments of length $ n_{\cE} $ and each of them is compressed separately. If there is a remainder smaller than $ n_{\cE} $, it is not compressed at all. This way, a $ \bfv $-avoiding sequence of length $n$ can be compressed to a uniquely decodable sequence of length $ n-\lfloor n/n_{\cE} \rfloor $. We abuse the notation $ \cE_{\bfv} $ to denote this generalized compressor as well. Similarly, let $\cD_\bfv$ denote the matching decoder of $\cE_{\bfv}$ for any $ \bfv \in \Sigma^{\ell} $.
%

Algorithm~\ref{alg:enc-subs-present} receives as an input $ \bfw $, a  sequence of length~{$n-1$} and outputs $ \bfx $, an $\ell$-tuples covering sequence of length $ n $.  
The goal of the algorithm is to shorten the input sequence enough in order to make room for appending a de Bruijn sequence of order $\ell$ at its end. The shortening procedure uses the family of compressors $\{ \cE_{ \bfv} \mid \bfv \in \Sigma^{\ell} \} $, based on the observation that as long as the sequence is not an $\ell$-tuples covering sequence, then it is $ \bfv $-avoiding for some tuple $ \bfv \in \Sigma^{\ell} $. Let  $ \bfs $ denote a fixed de Bruijn sequence of length $ 2^{\ell} + \ell - 1 $; $ \bfs $ can be {produced} with time complexity of $ \cO(\ell 2^\ell) $, see~\cite{Fr75}. 
The algorithm first sets $ \bfx = 0 \circ \bfw $ in order to mark the start of the encoding process for the decoder. Then, as long as $ \bfx $ is not $\ell$-tuples covering, the algorithm repeatedly shortens $ \bfx $ by finding an $\ell$-tuple $ \bfv $ that does not appear in $ \bfx $. The algorithm encodes the occurrence and compresses $\bfx $ using $ \cE_{\bfv} $. This process ends when $ \bfx $ is either an $\ell$-tuples covering sequence or it is short enough to be appended by $ \bfs $. Either way, this results with an   $\ell$-tuples covering sequence which is returned after being padded to length~$ n $.

\begin{algorithm}
	\caption{$ \ell $-tuples covering sequences  encoding}\label{alg:enc-subs-present}
	\algorithmicrequire{ A sequence $ \bfw \in \Sigma^{n-1} $} \\
	\algorithmicensure { A sequence $ \bfx \in \cR(n,\ell) $}
	\begin{algorithmic}[1]				
		\State{Set $ \bfx = 0 \circ \bfw $}
		\While{$ \bfx $ is not $\ell$-tuples covering and $ |\bfx| > n-|\bfs| $}
		\State{Pick $ \bfv \in \Sigma^{\ell} \setminus \{ \bfx_{i,\ell} : i \in [|\bfx| -\ell + 1] \} $}
		\State{Set $ \bfx = 1 \circ \bfv \circ  \cE_{\bfv}(\bfx)  $} \label{step:dec-x}
		\EndWhile
		\State{Return $\text{Pref}_n(\bfx \circ \bfs \circ 1^n) $} \label{step:ret-x}
	\end{algorithmic}
\end{algorithm}

We prove the correctness of Algorithm~\ref{alg:enc-subs-present} in the next two lemmas. 
\begin{lemma}\label{clm:dec-x}
	At every iteration of the while loop of Algorithm~\ref{alg:enc-subs-present}, at Step~\ref{step:dec-x} the size of $ \bfx $ decreases. 
\end{lemma}
\begin{IEEEproof}
	Let $ \bfx' $ denote the sequence $ \bfx $ after the execution of Step~\ref{step:dec-x} at some iteration of the loop. 
	Hence, 
	\begin{align*}
		|\bfx'| &= |\bfx| - \floorenv{\frac{|\bfx|}{n_{\cE}}} + \ell + 1
		\\&\overset{(a)}{\le} |\bfx| - \floorenv{\frac{n-|\bfs|}{n_{\cE}}} + \ell + 1
		\\&\overset{(b)}{\le} |\bfx| - \frac{n}{n_{\cE}} + \ell + 2
		\\&\overset{(c)}{\le} |\bfx| - \log n + \ell + 2 
		\\&\overset{(d)}{\le} |\bfx| - \log \log n - 2
		\\& < |\bfx|
	\end{align*}
	where (a) follows from the fact that throughout the while loop $ |\bfx| > n-|\bfs| $, (b) follows from $ |\bfs| < n_{\cE} $, and (c) follows from plugging $n_{\cE} = \frac{n}{\log n} $,  and (d) follows from plugging $ \ell \le \log n - \log \log n - 6$.	
\end{IEEEproof}

\begin{theorem}\label{lem:alg-subs-present}
	Algorithm~\ref{alg:enc-subs-present} successfully outputs a uniquely decodable  $\ell$-tuples covering sequence of length $n$. The time complexity of the algorithm and its decoder is $ \cO \left(\frac{n^2}{\log n \log \log n}\right) $. 
\end{theorem}
\begin{IEEEproof}
	Following Lemma~\ref{clm:dec-x}, the while loop terminates and the algorithm reaches Step~\ref{step:ret-x}. At Step~\ref{step:ret-x} $ \bfx $ satisfies at least one of the following, 1. $ |\bfx| \le n $ and $ \bfx $ is $\ell$-tuples covering, or 2. $ |\bfx| \le n-|\bfs| $. In both cases, $\text{Pref}_n(\bfx \circ \bfs \circ 1^n) $ is a  sequence of length $ n $ that contains an entire $\ell$-tuples covering sequence, $ \bfx $ in the first case and $ \bfs $ in the second case.
	
	As for the time complexity, at every iteration the algorithm invokes $\cE_{\bfv}$ at most $ \lfloor \frac{n}{n_{\cE}} \rfloor $ times, each requires $ \cO(\ell) $ from Theorem~\ref{th:alg1}. Additionally at every iteration finding $\bfv $ that is not contained in $\bfx$ can be done in $\cO(n)$. From Lemma~\ref{clm:dec-x}, $\bfx$ is shortened by $ \Theta(\log \log n) $ at every iteration, hence we have at most $ \cO\left(\frac{2^{\ell}}{\log \log n}\right)  =  \cO \left(\frac{n^2}{\log n \log \log n}\right) $ iterations, and we receive the result in the theorem statement. Note that the time complexity of constructing the de Bruijn sequence $\bfs$  is $\cO(\ell 2^\ell) = \cO(n) $ which does not affect the time complexity of the encoder.

In order to decode $ \bfw \in \Sigma^{n-1} $ given $ \bfx $, an output of Algorithm~\ref{alg:enc-subs-present}, we iteratively inverse the operation of the while loop using the set of decoders  $\{ \cD_{ \bfv} \mid \bfv \in \Sigma^{\ell} \} $. As long as $ x_0 = 1 $, we repeatedly extract $ \bfv = \bfx_{1,\ell} $ and decode the rest of $ \bfx $ using $ \cD_{\bfv} $. This process ends when $ x_0 = 0 $, where the decoder returns $ \bfw = \text{Pref}_{n-1}(\bfx) $. 

Note that as a result of the possible concatenation of $ \bfs $ and~$ 1^n $ to $ \bfx $ at Step~\ref{step:ret-x}, in some cases the values of $ \bfx $ in the matching iterations of the encoder and the decoder are not equal. In those cases, the decoded sequence contains an additional suffix and $ \cD_{\bfv} $ can be invoked on segments that were not outputs of $\cE_{\bfv}$ in Algorithm~\ref{alg:enc-subs-present}. However, this does not impact the correctness of the decoder since those segments will be trimmed from $ \bfx $ at the end of the decoding process.
\end{IEEEproof}

\begin{remark}
	Algorithm~\ref{alg:enc-subs-present} can be generalized to any $q > 2$ using a $q$-ary generalization of the compressor $ \cE_\bfv $. In this case, the algorithm can encode $q$-ary \ltuples covering sequences of length $ \ell \le \log_q n - \log_q \log_q n - 4 $ for $n$ large enough. 
\end{remark}

\subsection{{Upper Bound on the Rate Using de Bruijn Graph}}


In this section we use an enumeration technique which was first used to enumerate de Bruijn sequences using the de Bruijn graph~\cite{Mow66}. It was recently used to enumerate another generalization of de Bruijn sequences~\cite{MarEtzYaa21}. In this paper, this technique is used to derive an upper bound on the cardinality of $ \ell $-tuple covering sequences. For simplicity, we focus on binary sequences, although this technique can be extended to any alphabet of finite size.

Let $ n = 2^\ell + \ell - 1 + k $, for $k = f(n)$. For every selection of $ k $ $\ell$-tuples (with repetitions) we construct a graph $ G $ which is a generalization of the de Bruijn graph $ G_{\ell} $. 
The vertices of each graph are the same vertices of  $ G_{\ell} $ which are represented by the binary $(\ell-1)$-tuples.
The edges are the edges of $ G_{\ell} $ with additional parallel edges corresponding to each of the $ k $ $\ell$-tuples picked. There are $\binom{2^{\ell} + k -1}{k}$ different graphs which are generated this way.

A \emph{reverse spanning tree} $ T $ of a generalized graph $ G $ is a graph whose underlying graph is a tree rooted at some $ {\bfr \in \Sigma^{\ell-1}} $, it contains all the vertices of $ G $, and there is a unique directed path from each vertex $ v $ of $ T $ to $ \bfr $. Clearly, all the generalized graphs share the same set of reverse spanning trees as $ G_{\ell} $, and there are  $2^{2^{\ell-1}-\ell}$ such trees for each $ \bfr \in \Sigma^{\ell-1} $~\cite{Mow66}. 

For each graph $ G $, reverse spanning tree $ T $ and root vertex~$ \bfr $, we use a nondeterministic algorithm to attempt traversing all the edges of $ G $ exactly once, starting from the root vertex~$\bfr$. 
In order to ensure the uniqueness of the path with respect to~$T$, its edges are notated as \emph{starred}, and the algorithm leaves a vertex $v$ on a starred edge only if it is the last outgoing edge of~$v$ that was not traversed before (this algorithm is defined formally in~\cite{MarEtzYaa21}). If the algorithm succeeded traversing all the edges of~$G$, the result is a path that corresponds to an $ \ell $-tuples covering sequence of length $ n $. Notice that for some graphs, the algorithm might fail to traverse all the edges and to generate sequences. However, all the sequences of $\cR(n,\ell)$ are produced by this method and thus enumerating the paths constructed by such algorithm derives an upper bound for $ r(n,\ell) $. 

\begin{lemma}\label{lem:tree}
	For each reverse spanning tree, at most \break $ {2^{k + \ell}\dbinom{2^{\ell} + k -1}{k} } $ distinct acyclic sequences are constructed by the algorithm.
\end{lemma}
\begin{IEEEproof}
	First, there are $ \binom{2^{\ell} + t -1}{t} $ graphs and $ 2^{\ell-1} $ vertices in each graph. For each graph, root vertex and spanning tree, we can bound the number of different orders to traverse the edges by $ 2^{t+1} $. That is since each vertex besides the root that has $ t' \ge 0 $ parallel outgoing edges has at most $ 2^{t'} $ different options to traverse its outgoing edges, since the edges of $ T $ are traversed last. The root $\bfr$ does not share outgoing edges with $ T $ and therefore has at most $ 2^{t'+1} $ different orders. 
	Hence, we receive the value mentioned in the lemma statement.
\end{IEEEproof}

\begin{corollary}\label{cor:upper-bound-M}
	The total number of distinct acyclic sequences of length $ n $ formed by the algorithm is at most $ 2^{2^{\ell-1}+t} \binom{2^{\ell} + t -1}{t}$.
\end{corollary}

The value presented in Corollary~\ref{cor:upper-bound-M} provides an upper bound on  $ r(n,\ell) $.
Next, we derive an upper bound for the asymptotic rate of $ \ell $-tuples covering sequences, for different values of $ \ell $. Recall the following result of Lemma~\ref{clm:rate-lower}, applied for $ q=2 $,
\begin{corollary}\label{clm:rate-lower2}
	Let $ n = 2^\ell + \ell - 1 + f(n) $ for  $ \ell \in \mathbb{N} $. Then,
	$$ \mathbb{R}(\ell) \ge \begin{cases}
		\frac{1}{2} & f(n) = o(2^{\ell})\\
		\frac{2\alpha + 1}{2\alpha+2} & f(n) = \alpha 2^{\ell} + o(2^{\ell}) \text{ for constant } \alpha  > 0 \\
		1 & f(n) = \omega(2^{\ell})\\
	\end{cases}  $$
\end{corollary}

\begin{theorem}\label{th:rate-1/2}
	If $ n = 2^{\ell} + \ell  - 1 + f(n) $ where $ f(n) = o(2^{\ell}) $, then the asymptotic rate of $ \ell $-tuples covering sequences satisfies
	$$ \mathbb{R}(\ell) = \frac{1}{2}. $$
\end{theorem}
\begin{IEEEproof}
	Following Corollary~\ref{clm:rate-lower2} it is necessary to show that $ \mathbb{R}(\ell) \le 1/2 $. Let $ t = f(n) = o(2^{\ell}) $, then based on $ \binom{a}{b} \le 2^{aH_2(b/a)} $, where $ H_2 $ is the binary entropy function (see, e.g., \cite[Ch.10,~Sec.11,~Lem.7]{MacSlo78}), it is possible to derive that $ \log \binom{2^{\ell} + t -1}{t} = o(2^{\ell}) $. Hence,
	\begin{align*}
	\mathbb{R}(\ell) &\le \limsupup{\ell} \frac{ 2^{\ell-1}+t + \log \binom{2^{\ell} + t -1}{t} }{2^{\ell}} \\&= \frac{1}{2} + \limsupup{\ell}\frac{o(2^{\ell})}{2^{\ell}} = \frac{1}{2}.
	\end{align*}
\end{IEEEproof}

Similarly, when $ t = \alpha 2^{\ell} + o(2^{\ell}) $ for $ \alpha > 0 $, we have the next corollary which uses similar considerations as Theorem~\ref{th:rate-1/2}. 

\begin{corollary}\label{cor:rate-middle}
	Let $ n = 2^{\ell} + \ell - 1 + f(n) $ where $ f(n) = \alpha 2^{\ell} + o(2^{\ell}) $ for $ \alpha > 0 $, then the asymptotic rate of $ \ell $-covering sequences satisfies 
	$$ \mathbb{R}(\ell) \le  H\left(\frac{\alpha}{\alpha+1}\right) + \frac{2\alpha + 1}{2\alpha + 2}. $$
\end{corollary}
Note that the result of {Corollary~\ref{cor:rate-middle}} is useful only for small values of $ \alpha $ in the range $ 0 <\alpha < 1 $. Other values of $ \alpha $ are subject for future research.

Table~\ref{tab:compare} summarizes the results presented in this paper regarding the asymptotic rate of $ \ell $-tuples covering sequences.
\vspace{-3ex}
\begin{table}[h]
	\caption{Asymptotic rate of binary $ \ell $- tuples covering sequences $  n = 2^\ell + \ell - 1 + f(n) $}
	\vspace{-3ex}
	\begin{center}	\vspace{-3ex}	
		\label{tab:compare}	\begin{tabular}{lc}
			\hline\hline
			Case &  Result \\
			\hline
			$ f(n) = o(2^{\ell})$ & $\mathbb{R}(\ell) = 1/2 $ \\
			\hline
			$ f(n) =\alpha 2^{\ell} + o(2^{\ell}) \text{ ,  } \alpha  > 0 $ & $ \frac{2\alpha + 1}{2\alpha+2} \le \mathbb{R}(\ell) \le \min \left\{  H\left(\frac{\alpha}{\alpha+1}\right) + \frac{2\alpha + 1}{2\alpha+2},1 \right\} $ 		\\
			\hline
			$ f(n) = \omega(2^{\ell}) $ & $\mathbb{R}(\ell) = 1 $ \\
			\hline
		\end{tabular}\vspace{-2.5ex}
	\end{center}\vspace{-1.5ex}
\end{table}

%

	\section*{Acknowledgment}
S. Marcovich and E. Yaakobi were supported by the United States-Israel BSF grant no. 2018048. T. Etzion was supported by ISF grant no. 222/19.

\bibliographystyle{IEEEtranS}
\bibliography{mybib}
	
\end{document}